\documentclass[11pt]{article}

\usepackage[latin1]{inputenc}
\usepackage[american]{babel}
\usepackage[T1]{fontenc} 

\usepackage{fp, calc}
\usepackage{bm}

\usepackage{amsmath}
\usepackage{amssymb}
\usepackage{xspace}
\usepackage{theorem}
\usepackage{graphicx}
\usepackage{subfig}
\usepackage{epsfig}
\usepackage{ifpdf}
\usepackage{url,hyperref}
\usepackage{latexsym}
\usepackage{euscript}
\usepackage{xspace}
\usepackage{color}
\usepackage{makeidx}
\usepackage{picins,wrapfig}
\usepackage{stackrel}
\usepackage{amscd}
\usepackage[all]{xy}
\usepackage{multirow}
\usepackage{lineno}
\usepackage{algorithmic}
\usepackage{algorithm}

\usepackage{latexsym}
\usepackage{amsmath}
\usepackage{amssymb}
\usepackage{color}
\usepackage{graphicx}
\usepackage{algorithmic}
\usepackage{algorithm}

\newcommand{\norm}[1]{\| #1 \|}
\newcommand{\aff}{{\mbox{aff}}}

\newcommand{\defn}[1]{\emph{#1}}

\newcommand{\T}{\mathbb{T}} 

\newcommand{\e}{\varepsilon}

\newcommand{\Z}{\mathbb Z}
\newcommand{\R}{\mathbb R}

\newcommand{\conv}{{\rm {conv}}}

\newcommand{\link}{{\rm link}}
\newcommand{\str}{{\rm star}}

\newcommand{\vol}{{\rm vol}}
\newcommand{\vor}{{\rm Vor}}

\newcommand{\del}{{\rm Del}} 

\newcommand{\wit}{{\rm Wit}}

\newcommand{\rdela}{\del^{\alpha}}
\newcommand{\rdelo}{\del^{2\e}_0}

\newcommand{\sphere}[1]{\mathbb S^{#1}}
\newcommand{\pts}{L}

\newcommand{\probability}[1]{\mathsf{Pr}\left[ #1 \right]}

\newcommand{\twostr}{\mathrm{star}^2}

\newenvironment{proof}[1][{}]{
  \begin{trivlist}\item[]\textit{Proof #1}\quad}%
  {\hfill\hspace*{\fill}~$\square$\end{trivlist}}

\newtheorem{theorem}{Theorem}
\newtheorem{proposition}[theorem]{Proposition}

\newtheorem{lemma}[theorem]{Lemma}
\newtheorem{definition}[theorem]{Definition}

\usepackage{url,hyperref}

\usepackage{wrapfig}

\makeatother

\title{
A probabilistic approach to reducing the algebraic\\
complexity of computing Delaunay triangulations
}

\author{
Jean-Daniel Boissonnat
\footnote{
Geometrica, INRIA, Sophia Antipolis, France
\url{Jean-Daniel.Boissonnat@inria.fr}
}
\and
Ramsay Dyer
\footnote{
Johann Bernoulli Institute,
University of Groningen, 
Groningen, The Netherlands,
\url{r.h.dyer@rug.nl}
}
\and
Arijit Ghosh
\footnote{
D1: Algorithms \& Complexity,
Max-Planck-Institut f\"ur Informatik,
Germany
\url{agosh@mpi-inf.mpg.de}
}
}

\begin{document}

\maketitle

\pagenumbering{roman}
\thispagestyle{empty}


\begin{abstract}

Computing Delaunay triangulations in $\R^d$  
involves evaluating the so-called in\_sphere predicate that
determines if a point $x$ lies inside, on or outside the sphere
circumscribing $d+1$ points $p_0,\ldots ,p_d$. This predicate reduces to evaluating the
sign of a multivariate polynomial of degree $d+2$ in the coordinates
of the points $x, p_0,\ldots, p_d$. Despite much progress on exact
geometric computing, the fact that the degree of the polynomial
increases with $d$ makes the evaluation of  the sign of such a polynomial 
problematic except in very low dimensions. In this paper, we propose a
new approach that is based on the witness complex, a
  weak form of the Delaunay complex introduced by Carlsson and de
  Silva. 
The witness
  complex $\wit (L,W)$ is defined from two sets $L$ and $W$ in some
  metric space $X$: a finite set of points $L$ on which the complex
  is built, and a set $W$ of witnesses that serves as an approximation
  of $X$. A fundamental result of de Silva states that $\wit
  (L,W)=\del (L)$ if $W=X=\R^d$. In this paper, we give conditions on
  $L$ that ensure that the witness complex and the Delaunay
  triangulation coincide when $W$ is a finite set, and we
  introduce a new perturbation scheme to compute a perturbed set $L'$
  close to $L$ such that $\del (L')= \wit (L', W)$. Our perturbation algorithm is a geometric application of the Moser-Tardos 
constructive proof of  the Lov\'asz local lemma. 
  The only numerical operations we use are (squared)
  distance comparisons (i.e., predicates of degree 2).  The
  time-complexity of the algorithm   is sublinear in $|W|$.  Interestingly, although the algorithm does not
  compute any measure of simplex quality, a lower bound on the
  thickness of the output simplices can be guaranteed. 

\end{abstract}

\clearpage

\tableofcontents
\clearpage

\pagenumbering{arabic}

\section{Introduction}
\label{sec.introduction}

The witness complex was introduced by Carlsson and de Silva~\cite{cds-teuwc-04} 
as a weak form of the Delaunay complex that is suitable for finite metric spaces 
and is computed using only distance comparisons. The witness complex $\wit (L,W)$ 
is defined from two sets $L$ and $W$ in some metric space $X$: a finite set of 
points $L$ on which the complex is built, and a set $W$ of witnesses that serves 
as an approximation of $X$. A fundamental result of de Silva~\cite{vds-wdt-08} 
states that $\wit (L,W)=\del (L)$ if $W$ is the entire Euclidean space $X=\R^d$, 
and the result extends to spherical, hyperbolic and tree-like geometries. The 
result has also been extended to the case where $W=X$ is a smoothly embedded 
curve or surface of $\R^d$~\cite{aem-ww-2007}.  However, when the set $W$ of 
witnesses is  finite, the Delaunay triangulation and the witness complexes are 
different and it has been an open question to understand when the two structures 
are identical. In this paper, we answer this question and present an algorithm 
to compute a Delaunay triangulation using the witness complex.

We first give conditions on $L$ that ensure that the
witness complex and the Delaunay triangulation coincide when $W
\subset \R^d$ is a finite set (Section~\ref{sec:identitywitnessdelaunay}). Some 
of these conditions are purely
combinatorial and easy to check. In a second part (Section~\ref{sec:algo}), 
we show that those
conditions can be satisfied by slightly perturbing the input set $L$.
Our perturbation algorithm is a geometric application of the Moser-Tardos 
constructive proof of the general Lov\'asz
local lemma. Its analysis uses the notion of
protection of a Delaunay triangulation that we have previously introduced
to study the stability of Delaunay triangulations~\cite{bdg-stabdt-2014}. 
 

Our algorithm has several interesting properties and we believe that
it is a good candidate for implementation in higher dimensions.
\begin{description}
\item[1. Low algebraic degree.]  The only numerical operations used
  by the algorithm are (squared) distance comparisons (i.e.,
  predicates of degree 2).  In particular, we do not use orientation
  or in-sphere predicates, whose degree depends on the dimension $d$
  and are difficult to implement robustly in higher dimensions.


\item[2. Efficiency.] Our algorithm constructs the witness complex
  $\wit (L',W) = \del (L')$ of the perturbed set $L'$ in time
  sublinear in $|W|$. See Section~\ref{sec:compute.rdel}. 

\item[3. Simplex quality and Delaunay stability.]  Differently from all papers on this and
  related topics, we do not compute the volume or any measure of
  simplex quality. Nevertheless, through protection, a lower bound on
  the thickness of the output simplices can be guaranteed (see
  Theorem~\ref{thm:rdel.guarantees}), and the resulting Delaunay
  triangulation is stable with respect to small metric or point
  perturbations~\cite{bdg-stabdt-2014}.


\item[4. No need for coordinates.] We can construct Delaunay 
triangulations of points that come from 
some Euclidean space but whose actual positions are unknown. We simply 
need to know the 
interpoint distances.

\item[5. A thorough analysis.] Almost all papers in Computational Geometry rely on 
oracles  to evaluate predicates exactly and assume that the complexity of 
those oracles is $O(1)$. Our (probabilistic) analysis is more precise. We only 
use predicates of degree 2 (i.e. double precision) and the analysis fully covers 
the case of non generic data.
\end{description}

\subsection{Previous work}

Millman and Snoeyink~\cite{ms-cpvd-2010} developed  a degree-2 Voronoi diagram on a
$U\times U$ grid in the plane. The diagram  of $n$ points can
be computed using only double precision by a randomized incremental
construction in $O(n\log n\log U)$ expected time and $O(n)$ expected
space. The diagram also answers nearest neighbor queries, but
it doesn't  use sufficient precision to determine a Delaunay
triangulation.


Our work has some similarity with the $\e$-geometry introduced by
Salesin et al.~\cite{DBLP:conf/compgeom/SalesinSG89}.
For a given geometric problem, the basic idea is to compute the exact solution 
for a perturbed version of the input and to bound the size of this
implicit perturbation. The authors applied this paradigm to answering some geometric queries
such as deciding whether  a point lies inside a polygon. They do not
consider constructing geometric structures
like convex hulls, Voronoi diagrams or Delaunay triangulations. Also,
the required  precision is computed on-line and is not strictly
limited in advance as we do here.
 
Further developments have been undertaken under the 
name of controlled perturbation~\cite{dh-cp-2010}. The
purpose is again to actually perturb the input, thereby reducing the
required precision of the underlying arithmetic and avoiding explicit
treatment of degenerate cases.  A specific scheme for 
Delaunay
triangulations in arbitrary dimensions has been proposed by Funke et
al.~\cite{funke2005}. Their algorithm relies on a careful analysis of the usual
predicates of degree $d+2$ and is 
much more demanding than ours.

\subsection{Notation}

In the paper, $W$ and $L$ denote sets of points
in $\T ^d$ where $\T^d$ denotes 
the standard flat torus $\R^d/\Z^d$.  $L$ is finite but we will only
assume that $W$ is closed in $\T^d$. 
The points of $W$ are called the {\em witnesses} and the points of $L$
are called the {\em landmarks}.
To keep the exposition simpler, boundary issues will be handled in the
full version of the paper (see Section~\ref{app-boundary-issues}).


We say that $W \subset \T^d$ is an \defn{$\e$-sample} if for any $x \in \T^d$ there is a 
$w \in W$ with $\norm{w-x} < \e$. If $L \subset \T^d$ is a finite set, then it is a 
$\lambda'$-sample for $\T^d$ for all $\lambda'$ greater or equal to some finite $\lambda>0$.  
The parameter $\lambda$ is called the \defn{sampling radius} of $L$.  We further say 
that $L$ is \defn{$(\lambda,\bar{\mu})$-net} if $\norm{p-q} \geq \bar{\mu} \lambda$ 
for all $p,q \in L$. We call $\bar{\mu}$ the {\em sparsity ratio} of $L$. Note that, for any 
two points $p,q$ of a $(\lambda, \bar{\mu})$-net, we have 
$\bar{\mu}\lambda \leq \| p-q\| \leq 2\lambda$, which implies $\bar{\mu}\leq 2$.


A simplex $\sigma \subset L$ is a finite set.
We
always assume that $L$ contains a non-degenerate $d$-simplex, and we
demand that $\lambda \leq 1/4$. 
This bound on $\lambda$ 
avoids the topological complications associated with the periodic boundary conditions.

The volume of Euclidean $d$-ball with unit radius will be denoted by $U_{d}$.

%

\section{Delaunay and witness complexes}
\label{sec:witness}



\begin{definition}[Delaunay center and Delaunay complex]
A Delaunay center for a simplex $\sigma\subset L$ is a point $x \in \T^d$ that
satisfies
$$
  \| p-x\| \leq \| q-x\|, \;\;\forall p\in\sigma \;\; \mbox{and} \;\; \forall q\in L.
$$
The Delaunay complex $\del (L)$ of $L$ is the complex consisting of
all simplexes $\sigma \subset L$ that have  a Delaunay center.
\end{definition}
Note that $x$ is at equal distance from all the vertices of $\sigma$.
A Delaunay simplex is \defn{top dimensional} if is not the proper face of any 
Delaunay simplex. The affine hull of a top dimensional simplex has dimension $d$.
If $\sigma$ is top dimensional, the Delaunay center is the
circumcenter of $\sigma$ which we denote $c_{\sigma}$. We write
$R_{\sigma}$ for the circumradius of $\sigma$.



Delaunay~\cite{delaunay1934} showed that if the point set $L$ is
generic, i.e., 
if no empty sphere contains $d+2$ points on its boundary, then
$\del(L)$ is a 
triangulation of $\T^d$ (see the discussion in
Section~\ref{sec:identitywitnessdelaunay}), and 
any perturbation $L'$ of a finite set $L$ is generic with probability
1. We refer to 
this as \defn{Delaunay's theorem}.



We 
introduce now the witness complex that can be considered as a weak variant of the
Delaunay complex. 


\begin{definition}[Witness and witness complex]
  Let $\sigma$ be a   simplex with vertices in $L \subset \T^d$, and let $w$ be a point
  of $W \subseteq \T^d$. We say that $w$ is a \defn{witness of $\sigma$} if 
  $$
    \| w-p\| \leq \| w -q \|, \;\; \forall p\in \sigma  \;\; \mbox{and}
    \;\; \forall q\in L\setminus \sigma.
  $$
  The witness complex $\wit (L,W)$ is the complex consisting of all
  simplexes $\sigma$ such that for any simplex $\tau\subseteq
  \sigma$, $\tau$ has a witness in $W$.
\end{definition}

 \begin{figure}
   \begin{center}
     \includegraphics[width=0.50\linewidth]{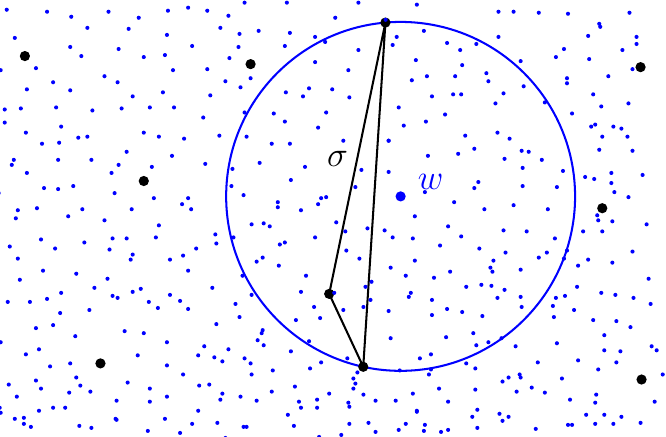} 
   \end{center}
   \caption{Blue points are the witnesses $W$ and the black points are 
   the landmarks $L$. The simplex $\sigma$ is {\em witnessed} by $w$.
   }
   \label{fig:witness.complex}
 \end{figure}

%
%

Observe that the only
predicates involved in the construction of $\wit (L,W)$ are (squared)
distance comparisons, i.e. polynomials of degree $2$ in the coordinates
of the points. This is to be
compared with the predicate that decides whether a point lies inside,
on or outside the sphere circumscribing a $d$-simplex which 
is a polynomial of degree $d+2$.


\section{Identity of witness and Delaunay complexes}
\label{sec:identitywitnessdelaunay}


In this section, we  
make the connection between Delaunay and witness complexes more precise.
We start with de Silva's result~\cite{vds-wdt-08}:
%
%
%
%
\begin{theorem}
$\wit (L, W)  \subseteq \del (L)$, and if $W= \T^d$ 
then $\wit (L, W) = \del (L)$.
\label{lemma-wit-in-del}
\end{theorem}

If $L$ is generic, we
know that $\del(L)$ is embedded in $\T^d$ by Delaunay's
theorem. 
It therefore follows from Theorem~\ref{lemma-wit-in-del} that the
same is true for $\wit (L,W )$. In particular, the dimension of $\wit
(L,W)$ is at most $d$. 




\subsection{Identity from protection}

When $W$ is not the entire space $\T^d$ but a finite set of points,
the equality between $\del(L)$ and $\wit(L,W)$
no longer holds.  
However, by requiring that the $d$-simplices of
$\del(L)$ be {\em $\delta$-protected}, a 
property introduced in \cite{bdg-stabdt-2014}, we are able to recover
the inclusion 
$\del(L) \subseteq \wit(L,W)$, and establish the equality between the
Delaunay 
complex and the witness complex with a discrete set of witnesses.

\begin{definition}[$\delta$-protection]
We say that a simplex $\sigma \subset L$ is $\delta$-protected  at 
$x\in \T^d$  if 
$$
\| x- q\| >  \| x-p\|  +\delta, \;\;
\forall p \in \sigma \;\; \mbox{and} \;\; \forall  q\in L\setminus \sigma.
$$
\end{definition}

We say that $\del (L)$ is \defn{$\delta$-protected} when each Delaunay
$d$-simplex of $\del (L)$ has a $\delta$-protected Delaunay center.
In this sense, $\delta$-protection is in fact a property of the point
set and we also say that $L$ is $\delta$-protected.  If $\del(L)$ is
$\delta$-protected for some unspecified $\delta>0$, we say that $L$ is
protected (equivalently $L$ is generic).  We
always assume $\delta < \lambda$ since it is impossible to have a
larger $\delta$ if $L$ is a $\lambda$-sample.  The following lemma is
proved in~\cite{2014arXiv1410.7012B}.
For a subcomplex $K \in \del (L)$, we define $\str(K ;\del(L))$ as the subcomplex consisting of all
the simplices that have at least one vertex in $K$ (note that
this definition departs from common usage). 
We will  use $\twostr(p)$ as a shorthand for $\str(\str(p;
\del(L)); \del(L))$. 


\begin{lemma}[Inheritance of protection]
\label{lemma-protection}
\label{lem:protection-inheritance}
Let $L$ be a $(\lambda,\bar{\mu})$-net 
and suppose $p \in L$. If every 
$d$-simplex in $\twostr(p)$ is $\delta$-protected,
then all simplices in $\str(p; \del(L))$ are at least
$\delta'$-protected where $\delta'=
\frac{\bar{\mu}\delta}{4d}$. Specifically, any Delaunay $k$-simplex  $\tau$
is $\delta'$-protected at a point $z_{\tau}$ which is the barycenter of the
circumcenters of a subset of $d-k+1$ $d$-cofaces of $\tau$. 
\end{lemma}



The following lemma is an easy consequence of the previous one.
%
\begin{lemma}[Identity from protection]
Let $L$ be a $(\lambda,\bar{\mu})$-net with $p \in L$.
If all the 
$d$-simplices in $\twostr(p)$ are $\delta$-protected 
and $W$ is an   $\e$-sample for $\T^d$ with
$\delta\geq \frac{8d\e}{\bar{\mu}}$, then 
$\str(p ; \wit (L,W))= \str(p; \del (L))$.
\label{lemma-wit=del}
\end{lemma}

\begin{proof}
By Theorem~\ref{lemma-wit-in-del}, we have  
$$
  \str(p; \wit (L,W)) \subseteq \str(p;\del (L)).
$$
We now prove
the other inclusion.  Let $\sigma$ be a simplex in $\str(p; \del (L))$ and set
$$
  \delta' = \frac{\bar{\mu}\delta}{4d} \geq 2\e.
$$

Let $\tau \subseteq \sigma$ be a face of $\sigma$. Then, by
Lemma~\ref{lem:protection-inheritance}, $\tau$ is $\delta'$-protected at a point $z_{\tau}$
such that
\begin{enumerate}
\item  $\| z_{\tau}-p_i\| = \| z_{\tau}-p_j\| = r\;\;\; \forall p_i,p_j\in \tau$
\item $\| z_{\tau}-p_l\| > r+\delta' \;\;\; \forall p_l\in L\setminus \tau $
\end{enumerate}

For any $x\in B(z_{\tau}, \delta'/2 )$, 
any $p_i\in \tau$  and any $p_l \in L\setminus\tau$, we have
\begin{eqnarray*}
  \|x-p_i\| \leq \| z_{\tau}-p_i\| + \| z_{\tau}-x\|
  \leq r + \frac{\delta'}{2}
\end{eqnarray*}
and
\begin{eqnarray*}
  \| x-p_l \| \geq \| z_{\tau}-p_l \| - \| x-z_{\tau} \|
  > r+\delta' -\frac{\delta'}{2}
  = r + \frac{\delta'}{2}.
\end{eqnarray*}
 
Hence, $x$ is a witness of $\tau$. Since $\e \leq
\delta'/2$, there must be a point  $w \in W$ in $B(z_{\tau}, \delta'/2 )$ which
witnesses $\tau$. Since this is true for all faces $\tau \subseteq
\sigma$, we have $\sigma \in \str(p; \wit(L,W))$.
\end{proof}


%
%

We end this subsection with a result proved 
in~\cite[Lemma 3.13]{bdg-stabdt-2014} that will be
useful in Section~\ref{sec:compute.rdel}.
For any vertex $p$ of a simplex $\sigma$, the {\em face oppposite} $p$ is the
face determined by the other vertices of $\sigma$, and is denoted by $\sigma_p$. 
The {\em altitude} of $p$ in $\sigma$ is the distance $D(p,\sigma) =
d(p, \aff (\sigma_p))$ from $p$ to the affine hull of $\sigma_p$. 
The altitude $D(\sigma )$ of $\sigma$ is
the minimum over all vertices $p$ of $\sigma$ of $D(p, \sigma)$. 
A poorly-shaped simplex can be
characterized by the existence of a relatively small altitude. The
{\em thickness} of a $j$-simplex $\sigma$ is the dimensionless
quantity $\Theta (\sigma)$ that evaluates to $1$ if $j=0$ and to
$    \frac{D(\sigma)}{j
      \Delta(\sigma)}$ otherwise, 
where $\Delta (\sigma)$ denotes the  {\em diameter} of $\sigma$, 
i.e. the length of its longest edge. 


\begin{lemma}[Thickness from protection]
  \label{thm:prot.thick}
  Suppose $\sigma \in \del(L)$ is a $d$-simplex with circumradius less than 
  $\lambda$ and shortest edge length greater than or equal to $\bar{\mu}\lambda$. 
  If every $(d-1)$-face of $\sigma$ is also a face of a $\delta$-protected $d$-simplex 
  different from $\sigma$, then the thickness of $\sigma$ satisfies 
  $$
    \Theta (\sigma) \geq \frac{\bar{\delta} \, (\bar{\mu}+\bar{\delta})}{8d}.
  $$

  In particular, suppose $p \in L$, where $L$ is a $(\lambda,\bar{\mu})$-net, and 
  every $d$-simplex in $\twostr(p)$ is $\delta$-protected, then every $d$-simplex in 
  $\str(p)$ is 
  $\left(\frac{\bar{\delta} \, \bar{\mu}}{8d}\right)$-thick.
\end{lemma}

\subsection{A combinatorial criterion for identity}
\label{sec:good.links}

The previous result will be useful in our analysis but does not help to compute $\del (L)$ from
$\wit (L,W)$ since the $\delta$-protection assumption requires
knowledge of $\del (L)$.  A more useful result in this context will be
given in
Lemma~\ref{lemma-wit=del-good-linksS}. Before stating the lemma, we
need to introduce some terminology and, in particular, the notion of
{\em good links}.


\begin{figure}  
  \begin{center}
    \includegraphics[width=0.350\textwidth]{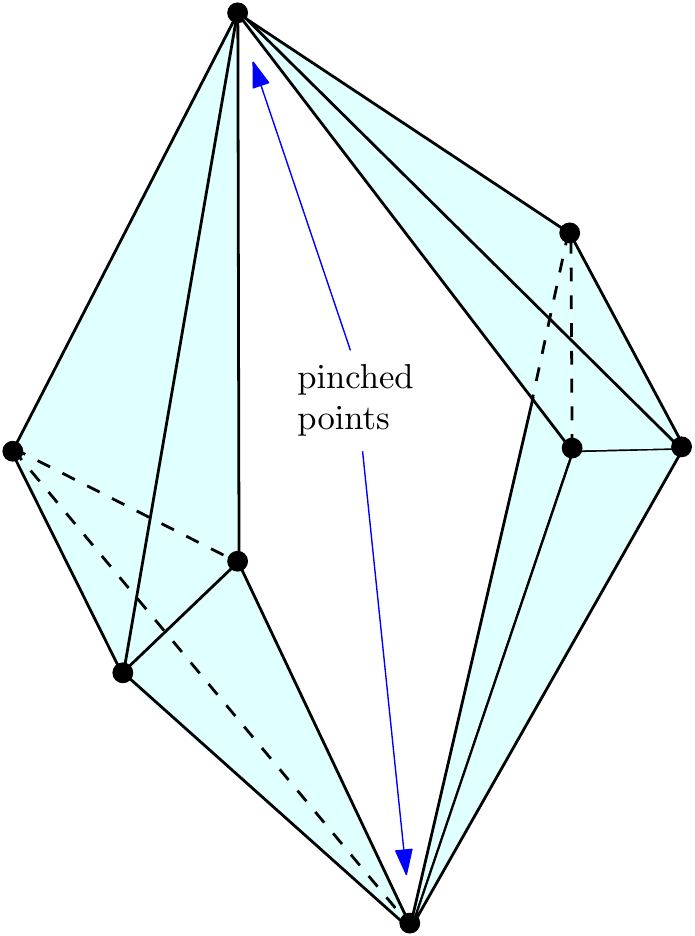}
  \end{center}
  \caption{Pinched simplicial torus is an example of 
  $2$-pseudo-manifold.}
  \label{fig:pseudo.manifold}
\end{figure}

A complex $K$ is a 
 \defn{$k$-pseudo-manifold} if it is a pure $k$-complex and every
$(k-1)$-simplex is the 
face of exactly two $k$-simplices.

\begin{definition}[Good links]
  Let $K$ be a complex with vertex set $L \subset \T^d$. We say $p \in
  L$ has a \defn{good link} if $\link(p; K)$ is a 
  $(d-1)$-pseudo-manifold. If every $p \in L$ 
  has a good link, we say $K$ has 
\defn{good links}. 
\end{definition}
For our purposes, a simplicial complex $K$ is a \defn{triangulation of
  $\T^d$} if it is a $d$-manifold
\emph{embedded} in $\T^d$. We observe that a triangulation has good links.
\begin{lemma}[Pseudomanifold criterion]
  \label{lem:pseudoman.criterion}
  If $K$ is a triangulation of $\T^d$ and $J\subseteq K$ has the same vertex set, then $J=K$ if and only if $J$ has good links.
%
\end{lemma}

\begin{proof}
  Let $L$ be the common vertex set of $K$ and $J$, and let $p$ be a point in $L$. 
  Since $K$ is a triangulation, $\link(p;K)$ is a simplicial $(d-1)$-sphere, which is 
  a manifold.  A $(d-1)$-pseudo-manifold $C_J$ cannot be properly contained in a connected 
  $(d-1)$-manifold simplicial complex $C_K$, because a $(d-1)$-simplex that does not 
  belong to $C_J$ cannot share a $(d-2)$-face with a simplex of $C_J$: Every $(d-2)$-simplex 
  of $C_J$ is already the face of two $(d-1)$-simplices of $C_J$. Since any two 
  $(d-1)$-simplices in $C_K$ can be connected by a path confined to the relative interiors 
  of $(d-1)$ and $(d-2)$-simplices, it follows that a $(d-1)$-simplex $\tau \in C_K$ must 
  also belong to $C_J$.  For an arbitrary simplex $\sigma \in C_K$, we have that $\sigma$ is 
  a face of a $(d-1)$-simplex $\tau \in C_K$, because $C_K$ is a pure $(d-1)$-complex 
  (it is a manifold). It follows that $\tau \in C_J$ and therefore $\sigma \in C_J$.

  Therefore, since $\link(p;J) \subseteq \link(p;K)$, if $J$ has good links we must have 
  $\link(p;J) = \link(p;K)$. It follows that $\str(p;J) = \str(p;K)$ for every $p \in L$, and 
  therefore $J=K$. If $J$ does not have good links it is clearly not equal to $K$, which does.
\end{proof}

We can now state the lemma 
that is at the heart of our algorithm. It follows from Theorem~\ref{lemma-wit-in-del}, 
Lemma~\ref{lem:pseudoman.criterion}, and Delaunay's theorem:
\begin{lemma}[Identity from good links]
  If 
  $L$ is generic and the vertices of $\wit (L,W)$ have good links, then $\wit (L,W)= \del(L)$.
\label{lemma-wit=del-good-linksS}
\end{lemma}


\section{Turning witness complexes into Delaunay complexes}
\label{sec:algo}



\label{sec:compute-del-via-wit}



Let, as before, 
$L$ be a finite set of landmarks and $W$ a finite set of witnesses.
In this section, we intend to use
Lemma~\ref{lemma-wit=del-good-linksS} to construct $\del(L')$, where
$L'$ is close to $L$, using only comparisons of (squared) distances.
The idea is to first construct the witness complex $\wit (L,W)$ which
is a subcomplex of $\del (L)$ (Theorem~\ref{lemma-wit-in-del}) that
can be computed using only distance comparisons. We then check whether
$\wit (L,W)=\del (L)$ using the pseudomanifold criterion
(Lemma~\ref{lem:pseudoman.criterion}).  While there is a  
vertex $p$ of $\wit (L,W)$ that has a bad link (i.e. a link that is
not a pseudomanifold), we perturb $p'$ and the set of vertices $I(p')$,
to be exactly defined in Section~\ref{sec:lll.analysis} (See Eq.~\ref{eq:def-I(p)}) 
that are responsible for 
the bad link $L(p') =\link(p', \wit (L',W)$,
and recompute the witness complex for the perturbed
points. 
We write $L'$ for the set of perturbed points at some stage of the algorithm. 
Each point $p'$ is randomly and independently taken from the so-called 
{\em picking ball} $B(p,\rho)$. Upon termination, we have $\wit(L',W)=\del (L')$. 
The parameter
$\rho$, the radius of the picking balls,  must satisfy Eq.~(\ref{eq:condition-parameters-relax})
to be presented later.
The steps are described in more detail in
Algorithm~\ref{alg:DT-from-wit}. 
The analysis of the 
algorithm 
relies on the Moser-Tardos constructive proof of 
Lov\'asz local lemma.

\begin{algorithm}[ht]
  \caption{Delaunay triangulation from witness complex}
  \label{alg:DT-from-wit}
  \begin{algorithmic}
    \STATE{\bf Input:}\quad  $L$, $W$, $\rho$, $\lambda$, $\mu$
    \WHILE{a vertex $p'$ of $\wit(L',W)$ has a bad link}
    \STATE perturb $p'$ and the points in $I(p')$ (defined in Eq.~\ref{eq:def-I(p)})
    \STATE update $\wit (L',W)$
     \ENDWHILE
    \STATE{\bf Output:}\quad $\wit (L',W)=\del  (L')$
  \end{algorithmic}
\end{algorithm}

\subsection{Lov\'asz local lemma}
\label{sec:LLL}

The celebrated Lov\'asz local lemma is a powerful tool to prove the
existence of combinatorial objects~\cite{the-probabilistic-method-alon-spencer}. 
Let $\mathcal{A}$ be a finite
collection of ``bad'' events in some probability space. The lemma
shows that the probability that none of these events occur is positive
provided that the individual events occur with a bounded probability and there 
is limited dependence among them. 
\begin{lemma}[Lov\'asz local lemma]
\label{lem-LLL}
   Let ${\cal A}=\{ A_{1}, \,
    \dots, \, A_{N}\}$ be a finite set of events in some probability space. Suppose 
    that each event $A_{i}$ is independent of all  
    but at most $\Gamma$ of the other events $A_{j}$, 
    and that $\probability{A_{i}} \leq \varpi$ for all $1 \leq i \leq N$. 
    If 
     $$ 
      \varpi \leq \frac{1}{e(\Gamma+1)}
     $$
    ($e$ is the base of the natural logarithm), then 
    $$
      \probability{\bigwedge_{i=1}^{N} \neg A_i} >0.
    $$
\end{lemma}
Assume that the events depend on a finite set of mutually independent
variables in a probability space.  Moser and
Tardos~\cite{mt-cplll-2010} gave a
constructive proof of Lov\'asz lemma leading to a simple and natural
algorithm that checks whether some event $A\in {\cal A}$ is violated
and randomly picks new values for the random variables on which $A$
depends. We call this a resampling of the event $A$. Moser and Tardos
proved that this simple algorithm quickly terminates, providing an
assignment of the random variables that avoids all of the events in
$\mathcal{A}$. 
The expected total
number of resampling steps 
is at most $N/\Gamma$.

\subsection{Three geometric lemmas}


We will be using the following geometric lemmas.

We will need the following technical result to prove Lemma~\ref{lem:boundsonIandS}.
\begin{lemma}
\label{lem:diameter-witness}
  If $L$ is a $\lambda$-sample of $\T^d$, the circumradius $R_{\sigma}$ of any
  simplex $\sigma$ in $\wit (L,W)$ is less than $\lambda$. The witnesses
  of $\sigma$ are at a distance less than $\lambda$ from the closest vertex
  of $\sigma$ and at a distance less than $3\lambda$ from any vertex of $\sigma$.
\end{lemma}
\begin{proof}
  By Theorem~\ref{lemma-wit-in-del}, $\sigma$ is a simplex of
  $\del(L)$ and, 
  since $L$ is a $\lambda$-sample of $\T^d$, the
  circumradius $R_{\sigma}$ is less than $\lambda$. 
  Let $w$ be a witness of $\sigma$ and $q$ the vertex of $\sigma$
  closest to $w$. Since $L$ is a $\lambda$-sample, $\| w-q\| <
  \lambda$. The diameter of $\sigma$ is thus at most $2R_{\sigma}<
  2\lambda$, and $\|w-p\| < 3\lambda$ for any vertex
  $p\in \sigma$.
\end{proof}

\begin{lemma}
\label{lem:perturbed-net-parameters}
 If $L$ is a $(\lambda,\bar{\mu})$-net of $\T^d$ and
 $4\bar{\rho}<\bar{\mu} $, then $\pts '$ is an $(\lambda ', \bar{\mu}')$-net, where 
 $$
  \lambda ' = \lambda (1+\bar{\rho}) \;\;\mbox{and}\;\;
  \bar{\mu}' = \frac{\bar{\mu}-2\bar{\rho}}{1+\bar{\rho}} \geq \frac{\bar{\mu}}{3}.
 $$
\end{lemma}


 \begin{figure}
   \begin{center}
     \includegraphics[width=0.40\linewidth]{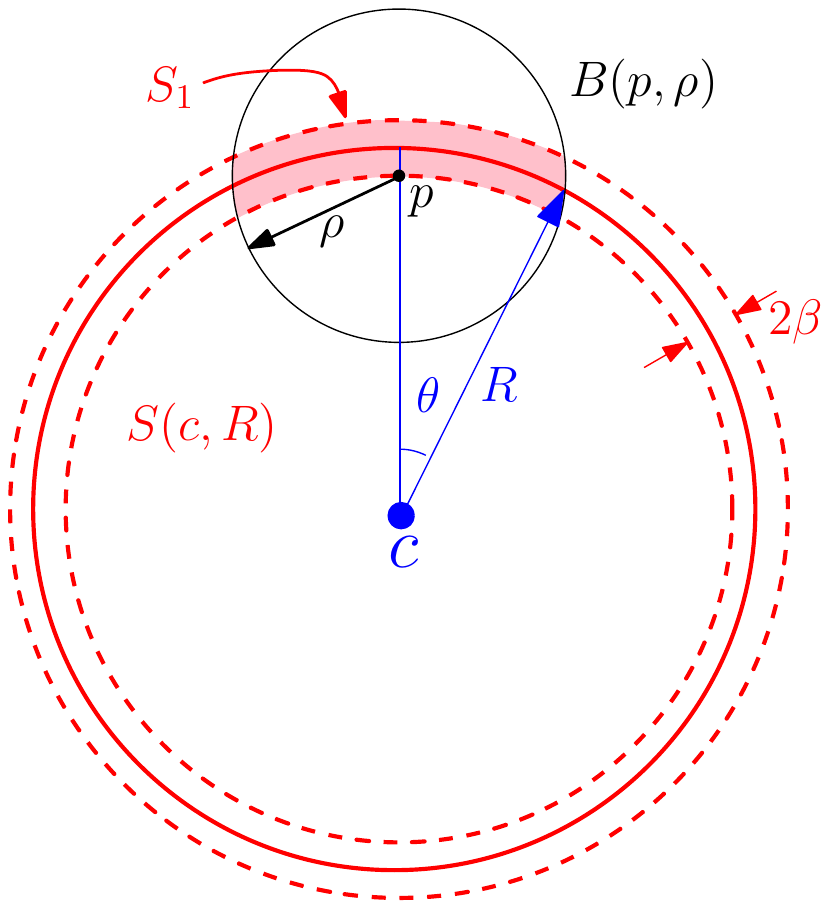} 
   \end{center}
   \caption{For Lemma~\ref{lem:shell.vol.bnd}.}
   \label{fig:forbidden.volume}
 \end{figure}

\begin{lemma}{\bf (See~\cite[Lemma~5.2]{boissonnat:hal-00806107})}
  Let $S(c,R)$ be a hypersphere of $\R^d$ of radius $R$ centered at $c$
  and $T_{\delta}$ the spherical shell
  $T_{\delta}=B(c,R+\delta)\setminus B(c,R)$. Let, in addition, $B_{\rho}$ denote any $d$-ball of
  radius $\rho<R$. We have
  $$
    \vol_d (T_{\delta}\cap B_{\rho}) \leq U_{d-1}  \left(
    \frac{\pi}{2}\rho \right)^{d-1} \delta,
  $$
  where $U_{d-1}$ denotes the volume of a Euclidean
  $(d-1)$-ball with unit radius.
%
  \label{lem:shell.vol.bnd} 
\end{lemma}



\subsection{Correctness of the algorithm}
\label{sec:lll.analysis}

We first bound the density radius $\lambda '$ and the sparsity
ratio $\bar{\mu}'$ of any perturbed point set $L'$.
We will write  $\rho= \bar{\rho}\lambda$ and
$\mu=\bar{\mu}\lambda$, and assume $\bar{\rho}< \bar{\mu}/4$.

We refer to the terminology of the  Lov\'{a}sz local lemma.
Our {\em variables} are the points of $L'$ which are randomly and
independently taken from the picking balls  $B(p, \rho)$, $p\in L$. 

The {\em events} are associated to points of $L'$, the 
vertices of $\wit (L',W)$. We say that an
event happens at $p'\in L$ when the link $L(p')$ of $p'$ in $\wit (L',W)$ is
not good, i.e., is not a pseudomanifold. 
We know from Lemma~\ref{lemma-wit=del}  
that if  $p'$ is a 
vertex of $\wit (L',W)$ and $L(p')$ is not good,
then there must exist a $d$-simplex in $\twostr(p')$ that is not
$\delta$-protected for $\delta = 8d\e/\bar{\mu}$.   
We will denote by 
\begin{itemize}
	\item
		$I_{1}(p') : $ the set of points  of $L'$ that can be
                in $\twostr(p')= \str(\str(p'; \del(L'); \del(L'))$
	
	\item
		$I_{2}(p') : $ the set of points of $L'$ that can
                violate the $\delta$-protected zone
                $Z_{\delta}(\sigma') =
                B(c_{\sigma},R_{\sigma}+\delta )\setminus  B(c_{\sigma},R_{\sigma})$
		for some $d$-simplex $\sigma'$ in  $\twostr(p')$

	\item
		$I(p') :$ Since computing $I_{1}(p')$ and $I_{2}(p')$ exactly is 
		difficult, we will rather compute the set
		\begin{equation}
		  I(p') = \left\{q'\; : \; q \in L \cap B(p, R)
		  \;\mbox{where}\; R = 5\lambda+\frac{3\bar{\mu}
                    \lambda}{2} \right\}.
\label{eq:def-I(p)}
              \end{equation}
              Using triangle inequality, 
		Lemma~\ref{lem:perturbed-net-parameters}, and the facts that $\bar{\rho} < \bar{\mu}/4$
		and $\delta \leq \lambda$, we can show that
		$$
		  I(p') \supseteq \tilde I_{1}(p') \cup \tilde I_{2}(p'),
		$$
                where $\tilde{I}_i(p')$ is the set of points in $L$ that correspond to points in $I_i(p') \subset L'$.
	\item
		$S(p') :$  the set of $d$-simplices with vertices in
		$I_{1}(p')$ that can belong to $\twostr(p')$ 
\end{itemize}
%
%
%
%
%
%
%
%
The probability $\varpi_1(p')$  that
$L(p')$ is not good is at most the probability $\varpi_2(p')$ that
one of the simplices of $S(p')$, say $\sigma'$,  has its
$\delta$-protecting zone $Z_{\delta}(\sigma')$ violated by some point of $L'$. Write $\varpi_3(q',\sigma')$ for
the probability that $q'$ belongs to the $\delta$-protection zone  of the $d$-simplex $\sigma'$.  We have 
\begin{equation}
 \varpi_1(p') \leq \varpi_2(p') \leq \sum_{q'\in I_{2}(p')}\, \sum_{\sigma'\in S(p')}
\varpi_3(q',\sigma') 
\label{eq:varpi123}
\end{equation}
The following lemmas
provide  upper bounds on 
$|I(p')|$, $|S(p')|$, 
$\Gamma$ and $\varpi_3(q',\sigma')$. 
\begin{lemma}
 $|I(p')| \leq  I= \left( \frac{14}{\bar{\mu}
  }\right) ^d $ and $|S(p')|  \leq K {=}
\frac{I ^{d+1}}{(d+1)!}$.
\label{lem:boundsonIandS}
\end{lemma}

\begin{proof} 
  Since any two points in $L$ are at
  least $\bar{\mu}\lambda$ apart, a volume argument then gives the
  announced bound. Referring to \eqref{eq:def-I(p)} we have
  $$
  |I(p')|
      \leq \left( \frac{5\lambda + \frac{3\bar{\mu}\lambda}{2}
          + \frac{\bar{\mu} \lambda}{2}}{\frac{\bar{\mu}\, \lambda}{2}}\right) ^d
    \leq \left( 4+\frac{10}{\bar{\mu}}\right) ^d
    \leq \left(\frac{14}{\bar{\mu}} \right)^{d}.
  $$ 

  We thus have 
  $$
	|S(p')| \leq \left(  \begin{array}{c} I \\ d +1  \end{array} \right) 
	\leq \frac{I ^{d+1}}{(d+1)!}.
  $$
\end{proof}

\begin{lemma}
  An event is independent of all but at most 
  $\Gamma=\left(\frac{27}{\bar{\mu}}\right)^d$ 
  other events.
  \label{lem:gamma-wit}
\end{lemma}
\begin{proof}
  Two events  $p'$ and $q'$ 
  are independent if $I(p')\cap
  I(q')= \emptyset$.  Referring to $R$ defined in \eqref{eq:def-I(p)},
  this implies $p'$ and $q'$ are independent if 
  $\| p - q\| < 2R = 10\lambda + 3 \bar{\mu}\lambda$. 
  Hence an upper bound on the  number of events
  that may  not be independent from the event at $p'$ is given by a volume argument as in
  the proof of Lemma~\ref{lem:boundsonIandS}. Here we count points since
  the events are associated to points:
  $$
     \left( \frac{10\lambda+ 3\bar{\mu}\lambda+
    \frac{\bar{\mu}\,\lambda}{2}}{\frac{\bar{\mu}\,
    \lambda}{2}}\right) ^d \leq \left( 7+\frac{20}{\bar{\mu}
    }\right)^d \leq \left(\frac{27}{\bar{\mu}}\right)^d \stackrel{{\rm
      def}}{=} \Gamma.
  $$
\end{proof}

\begin{lemma}
  $\varpi_3(q', \sigma')\leq  2\pi^{d-1} \frac{\delta}{\rho}$.
  \label{lem:varpi3}
\end{lemma}
\begin{proof}
  The volume of Euclidean $d$-ball with unit radius will be denoted by $U_{d}$.
  The probability $\varpi_3(q', \sigma') $ is the ratio between the
  volume of the portion of the picking region that is in the spherical
  shell $B(c_{\sigma'},R_{\sigma'} +\delta)\setminus B(c_{\sigma'}, R_{\sigma'})$
  and the volume of the picking region. 
  Hence, using~\cite[Lem.~5.2]{boissonnat:hal-00806107}
  and the crude 
  bound $\frac{U_{d-1}}{U_d}\leq 2^d$, we obtain
  $$
    \varpi_3(q', \sigma') \leq \frac{U_{d-1}\, (\frac{\pi}{2}\,
    \rho)^{d-1}\, \delta}{U_d\, \rho^d}  \leq   2\pi^{d-1} \,
    \frac{\delta}{\rho}.
  $$
\end{proof}
Using Eq.~\eqref{eq:varpi123}, and 
Lemmas~\ref{lem:boundsonIandS} and \ref{lem:varpi3}, we conclude that
$$ 
  \varpi_1(p') \leq 2\pi^{d-1} \, I\, K\, \frac{\delta}{\rho}.
$$

An event depends on at most $\Gamma$ other events. Hence, to apply
the Lov\'asz Local Lemma~\ref{lem-LLL}, it remains to ensure that
$\varpi_1(p')\leq \frac{1}{(e(\Gamma +1))}$. In addition, we also need that
$\delta \geq \frac{8d\e}{\bar{\mu}'}$ to be able to apply
Lemma~\ref{lemma-wit=del}. We thefore need to satisfy
the following inequality:
\begin{eqnarray}
\frac{8d\e }{\bar{\mu}'} \leq \delta \leq J \rho
\;\;\mbox{where}\;\;
J^{-1}\stackrel{{\rm def}}{=}  2 e \pi^{d-1} I K(\Gamma+1)
= \left(\frac{2}{\bar{\mu}}\right)^{O(d^{2})}
 \label{eq:condition-parameters-relax-1}
\end{eqnarray}
%
%
%
%
%
%
%
%
%
Observe that $I$, $K$, $\Gamma$ and $J$ depend only on $\bar{\mu}$ and
$d$.  
We conclude that the
conditions of  the Lov\'asz local lemma   hold
if the parameter  $\rho$  satisfies
\begin{equation}
\frac{\mu}{4}\geq \rho \geq 
\frac{24d\e}{\bar{\mu} J}
\label{eq:condition-parameters-relax}
\end{equation}

Hence, if $\e$ is sufficiently small, we can
fix $\rho$ so that  Eq.~(\ref{eq:condition-parameters-relax}) holds.  The
algorithm is then guaranteed to terminate. By Lemma~\ref{lemma-wit=del-good-linksS}, the output
is $\del (L')$.


It follows from Moser-Tardos theorem that the expected number of times 
the "while-loop" in Algorithm~\ref{alg:DT-from-wit} runs is 
$O\left( \frac{|L|}{\Gamma} \right)$ and since $|I(p')| \leq I$, we
get that the 
number of point perturbations
performed by Algorithm~\ref{alg:DT-from-wit} is 
$O\left(\frac{I\, |L|}{\Gamma}\right)= O(|L|)$ on expectation, where
the constant in the $O$ depends only on $d$ and {\em not} on the
sampling parameters $\lambda$ or $\mu$.
We sum up the results of this section  in
%
\begin{theorem}
\label{th:pert-LLL}
Under Eq.~(\ref{eq:condition-parameters-relax}), Algorithm~\ref{alg:DT-from-wit} 
terminates and outputs the Delaunay
triangulation of some set $L'$ whose distance to $L$ is at most
$\rho$. The number of point perturbations performed by the algorithm
is $O\left(|L|\right)$.
\end{theorem}


%

\section{Sublinear algorithm}
\label{sec:compute.rdel}

When the set $L'$ is generic, $K=\wit (L',W)$ is
embedded in $\T^d$ and is therefore $d$-dimensional.  It is well known
that the $d$-skeleton of $\wit (L',W)$ can be computed in time $O((|W|
+ |K|)\log |L'|)$ using only distance
comparisons~\cite{DBLP:journals/algorithmica/BoissonnatM14}. 
Although easy and general, this construction is not efficient when $W$
is large. 

In this section, we show how to implement an algorithm with execution time 
sublinear in $|W|$. 
We will assume that the points of $W$ are located at the centers of a grid, 
which is no real loss of generality. 
The idea is to restrict our attention to a subset of $W$, namely the set 
of \defn{full-leaf-points} introduced in Section~\ref{sec:eWcenters}.  These are points that 
may be close to the circumcentre of some $d$-simplex. A crucial observation is that if a 
$d$-simplex has a bounded thickness, then we can efficiently compute a bound 
on the number of its full-leaf-points.  This observation will also allow us to 
guarantee some protection (and therefore thickness) on the output simplices, as 
stated in Theorem~\ref{thm:rdel.guarantees} below.

The approach is based on the \defn{relaxed Delaunay complex}, which is related 
to the witness complex, and was also introduced by de Silva~\cite{vds-wdt-08}. 
We first introduce this, and the structural observations on which the algorithm 
is based.

\subsection{The relaxed Delaunay complex}
\label{sec:rdel.overview}

The basic idea used to get an algorithm sublinear in $|W|$ is to choose witnesses 
for $d$-simplices that are close to being circumcentres for these simplices. With 
this approach, we can in fact avoid looking for witnesses of the lower dimensional 
simplices. The complex that we will be computing is a subcomplex of a relaxed 
Delaunay complex:
\begin{definition}[Relaxed Delaunay complex]
Let $\sigma \subset L'$ be a simplex. An \defn{$\alpha$-center} of $\sigma$ is 
any point $x\in \T^d$ such that 
$$ 
  \| x -p\| \leq  \| x -q\| + \alpha, \; \; \forall \, p, \, q \in \sigma. 
$$
We say that $x$ is an \defn{$\alpha$-Delaunay center} of $\sigma$ if
$$
  \| x -p\| \leq  \| x -q\| + \alpha, \;\; \forall p \in \sigma 
  \; \mbox{and} \; \forall q \in L'. 
$$
The set of simplices that have an $\alpha$-Delaunay centre in $W$ is a simplicial 
complex, called the \defn{$\alpha$-relaxed Delaunay complex}, and is denoted 
$\rdela(L',W)$.
\end{definition}
We say $w \in W$ is an \defn{$\alpha$-witness} for $\sigma \subset L'$ if 
$$
  \|w-p\| \leq \|w-q\| + \alpha,\;\;  \forall p \in \sigma\;\; 
  \mbox{and}\;\;\forall q \in L'\setminus \sigma. 
$$
We observe that $w \in W$ is an $\alpha$-Delaunay center if and only if it is 
an $\alpha$-center and also an $\alpha$-witness.

\begin{lemma}
  \label{lem:rdel.props}
  The distance between an $\alpha$-Delaunay center for $\sigma \in \rdela(L',W)$ 
  and the farthest vertex in $\sigma$ is less than $\lambda' + \alpha$. In particular, 
  $\Delta_{\sigma} < 2\lambda' + 2\alpha$.

  If $\tau \in \del(L')$ and $c$ is a Delaunay center of $\tau$, then any point in 
  $B(c,r)$ is a $2r$-Delaunay centre for $\tau$. Thus $\del(L') \subseteq \del^{2\e}(L',W)$.
\end{lemma}

If, for some $\delta \geq 0$, all the $d$-simplices in $\rdela(L',W)$ have a $\delta$-protected
circumcentre, then we have that $\rdela(L',W) \subseteq \del(L')$, and
with $\alpha \geq 2\e$, it follows (Lemma~\ref{lem:rdel.props}) that
$\rdela(L',W) = \del(L')$, and $\del(L')$ is itself $\delta$-protected
and has good links. 

Reviewing the analysis of the Moser-Tardos algorithm of Section~\ref{sec:lll.analysis}, we observe that the exact same estimate of $\varpi_2(p')$ that serves as an upper bound on the probability that one of the simplices in $\twostr(p', \del (L'))$ is not $\delta$-protected at its circumcenter, also serves as an upper bound on the probability that one of the simplices in $\twostr(p', \rdela (L',\T^d))$ is not $\delta$-protected at its circumcenter, provided that $4\alpha + \delta \leq \lambda'$ (using the diameter bound of $2\lambda' + 2\alpha$ from Lemma~\ref{lem:rdel.props}), which we will assume from now on.  We can therefore modify Algorithm~\ref{alg:DT-from-wit} by replacing $\wit (L',W)$ by $\rdela(L',W)$.


We now describe how to improve  this algorithm to make it efficient. For our purposes it will be sufficient to set $\alpha=2\e$.
%
%
In order to obtain an algorithm sublinear in $|W|$, we will not
compute the full $\del^{2\e}(L',W)$  but only a subcomplex  we call
$\rdelo(L',W)$. The exact definition of $\rdelo(L',W)$ will be given
in Section~\ref{sec:fastalgo}, but the idea is  to only consider
$d$-simplices that show the properties of being $\Theta_0$-thick for
some parameter $\Theta_0$ to be defined later (Eq.~\ref{eq:defn.ThetaO} and \ref{eq:defn.delta}). This will allow us to restrict our attention to points of $W$ that
lie near the circumcentre. As explained in Section~\ref{sec:fastalgo},
this is done without explicitly computing thickness or circumcentres. 

As will be shown in Section~\ref{sec:sublin.correct} (Lemma~\ref{lem:rdel=del}),  the
modification of Algorithm~\ref{alg:DT-from-wit} that computes
$\rdelo(L',W)$ instead of $\wit(L',W)$ will terminate and output a complex
$\rdelo(L',W)$ with good links. However, this is not sufficient to
guarantee
that the
output is correct, i.e., that $\del^{2\e}(L',W)=\del (L')$. In order to obtain this guarantee, we
insert an extra procedure \texttt{check()}, which, without affecting
the termination guarantee, will ensure that the
simplices of $\rdelo(L',W)$ have $\delta^*$-protected circumcenters
for a positive $\delta^*$.  It follows that $\rdelo(L',W) \subseteq
\del(L')$ and, by 
Lemma~\ref{lem:pseudoman.criterion}, that $\rdelo(L',W) = \del(L')$. The modification of
Algorithm~\ref{alg:DT-from-wit} that we will use is outlined in
Algorithm~\ref{alg:rdel}. 
\begin{algorithm}[ht]
  \caption{Protected Delaunay triangulation from $\rdelo(L',W)$}
  \label{alg:rdel}
  \begin{algorithmic}
    \STATE{\bf input:}\quad  $L$, $W$, $\rho$, $\e$, $\lambda$, $\mu$
    \STATE $L' \gets L$
    \STATE {\bf compute:} $\rdelo(L',W)$
    \WHILE{a vertex $p'$ of $\rdelo(L',W)$ has a bad link or \texttt{check}$(p') = \mathrm{FALSE}$}
    \STATE perturb $p'$ and the points in $I(p')$
    \quad ($I(p')$ is defined in Section~\ref{sec:lll.analysis}.)
    \STATE update $\rdelo(L',W)$
     \ENDWHILE
    \STATE{\bf output:}\quad $\rdelo(L',W)=\del(L')$ with
    $\delta^*$ protection (See Lemma~\ref{lem:defn.delta.star})
  \end{algorithmic}
\end{algorithm}


We describe the details of computing $\rdelo(L',W)$ and of the
\texttt{check()} procedure in the following subsections.

\subsection{Computing relaxed Delaunay centers}
\label{sec:eWcenters}

We observe that the $\alpha$-Delaunay centers of a $d$-simplex $\sigma$ are close 
to the circumcenter of $\sigma$, provided that $\sigma$ has a bounded thickness:
\begin{lemma}[Clustered $\alpha$-Delaunay centers]
  \label{lem:a.del.ctrs}
  Assume that $L'$ is a $(\lambda', \bar{\mu}')$-sample.  Let $\sigma$
  be a non degenerate $d$-simplex, and let $x$ be an $\alpha$-center
  for $\sigma$ at distance at most $C\lambda'$ from the vertices of
  $\sigma$, for some constant $C>0$.  Then $x$ is at distance at most
  $\frac{C \alpha}{\Theta (\sigma) \bar{\mu}' }$ from the circumcenter $c_{\sigma}$ of $\sigma$. 
  In particular, if $x$ is an $\alpha$-Delaunay center for $\sigma$, then 
  $$
    \| c_{\sigma} - x\| < \frac{2\alpha}{\Theta (\sigma) \bar{\mu}'}.
  $$
\end{lemma}
\begin{proof}
  Using \cite[Lemma 4.3]{bdg-stabdt-2014} and $\Delta(\sigma)\geq
\bar{\mu}'\lambda'$, we get
$$ 
	\| w_{\sigma}-c_{\sigma}\| \leq \frac{C \lambda'
  	\alpha}{\Theta(\sigma)\Delta(\sigma)} 
	\leq \frac{C   \alpha}{\Theta(\sigma)\bar{\mu}' }.
$$
The second assertion follows from this together with Lemma~\ref{lem:rdel.props}, 
and the assumption that $\alpha < \lambda'$.
\end{proof}

%

It follows from Lemma~\ref{lem:a.del.ctrs} that $\alpha$-Delaunay centers are 
close to all the bisecting hyperplanes 
$$
  H_{pq} = \{x \in \R^d \,|\, \|x-p\| =\|x-q\|, \,\,p,q\in \sigma\}.
$$
The next simple lemma asserts a kind of qualitative converse:
\begin{lemma}
  \label{lem:grid-a-center}
  Let $\sigma$ be a $d$-simplex and $H_{pq}$ be the bisecting hyperplane of 
  $p$ and $q$. A point $x$ that satisfies $d(x,H_{pq}) \leq \alpha$, for any 
  $p,q\in \sigma$ is a $2\alpha$-center of $\sigma$.
\end{lemma}

\label{sec:computing-acenters}

Let $\sigma$ be a $d$-simplex of $\rdela(L',W)$ and let $\bar{\Omega}$ be the smallest box with edges parallel to the coordinate axes that contains $\sigma$.  Then the edges of $\bar{\Omega}$ have length at most $2\lambda' + 2\alpha$ (Lemma~\ref{lem:rdel.props}). Any $\alpha$-Delaunay center for $\sigma$ is at a distance at most $\lambda' + \alpha$ from $\bar{\Omega}$. Therefore all the $\alpha$-Delaunay centers for $\sigma$ lie in an axis-aligned hypercube $\Omega$ with the same center as $\bar{\Omega}$ and with side length at most $4\lambda'+4\alpha < 5\lambda'$.  Observe that the diameter (diagonal) $z$ of $\Omega$ is at most $5\lambda'\sqrt{d}$.


Our strategy is to first compute the $\alpha$-centers of $\sigma$ that belong to 
$\Omega \cap W$ and then to determine which ones are $\alpha$-witnesses for $\sigma$. 
Deciding if an $\alpha$-center is an $\alpha$-witness for $\sigma$ can be done in 
constant time since $L'$ is a $(\lambda', \bar{\mu}')$-net and  
$\Delta_{\sigma} \leq 2\lambda' + 2\alpha$ (Lemma~\ref{lem:rdel.props}).

We take $\alpha = 2\e$.  To compute the $2\e$-Delaunay centers of $\sigma$, we will 
use a pyramid data structure.  The pyramid consists of at most 
$\log \frac{z}{\e}$
levels. Each level $h>0$ is a grid of resolution 
$2^{-h}z$. 
The grid at level 
$0$ consists of the single cell, $\Omega$.  Each node of the pyramid is associated 
to a cell of a grid. The children of a node $\nu$ correspond to a subdivision into 
$2^d$ subcells (of the same size) of the cell associated to $\nu$. The leaves are 
associated to the cells of the finest grid whose cells have diameter $\e$.

A node of the pyramid that is intersected by all the bisecting hyperplanes of 
$\sigma$ will be called a \defn{full node} or, equivalently, a {\em full cell}.  
By our definition of $W$, a cell of the finest grid contains an element of $W$ 
at its centre. The \defn{full-leaf-points} are the elements of $W$ associated 
to full cells at the finest level. By Lemma~\ref{lem:grid-a-center}, the 
full-leaf-points are $2\e$-centers for $\sigma$. In order to identify the 
full-leaf cells, we traverse the full nodes of the pyramid starting from the 
root. Note that to decide if a cell is full, we only have to decide if two 
corners of a cell are on opposite sides of a bisecting hyperplane, which 
reduces to evaluating a polynomial of degree 2 in the input variables.

We will compute a bound $n_{\sigma}(\e)$ on the number  of full cells in the pyramid.  
Consider a level in the pyramid and the associated grid $G_{\omega}$ whose cells have 
diameter $\omega$.  Any point in a full cell of $G_{\omega}$ is at distance $\omega$ 
from the bisecting hyperplanes of $\sigma$ and so, by Lemma~\ref{lem:grid-a-center}, 
is a $2\omega$-center for $\sigma$.  By construction of $\Omega$, the distance between 
a point $w \in \Omega$ and a vertex $p \in \sigma$ is bounded by 
$$
	\|w-p\| \leq (3\lambda'+3\alpha)\sqrt{d} < 4\lambda'\sqrt{d}. 
$$
Therefore, it follows 
from Lemma~\ref{lem:a.del.ctrs} that all the full cells of $G_{\omega}$ are contained 
in a ball of radius $\frac{4\sqrt{d}\omega}{\Theta(\sigma)\bar{\mu}'}$ centered at 
$c_{\sigma}$. Letting $C$ be a hypercube with the dimensions of a cell in $G_{\omega}$, 
a volume argument gives a bound on $n_{\omega}$, the number of full cells in $G_{\omega}$:
\begin{equation*}
  n_{\omega} \leq \frac{\vol (B)}{\vol (C)}
  =
  U_d\left(\frac{4\sqrt{d}\omega }{\Theta(\sigma)\bar{\mu}'}\right)^d \left(\frac{\sqrt{d}}{\omega}\right)^d
  = U_d \left(\frac{4d}{\Theta(\sigma)\bar{\mu}'} \right)^d,
\end{equation*}
where $U_d$ is the volume of the Euclidean ball of unit radius. 

Observe that $n_{\omega}$ does not depend on $\omega$. Hence we can apply
this bound to all levels of the pyramid, from the root level
to the leaves level. Since there are
at most $\log\frac{5\lambda'\sqrt{d}}{\e}$ levels, 
we conclude with the following lemma:
\begin{lemma}
  The number of full cells is less than 
  $$
    n_{\sigma}(\e) = \frac{U_d(4d)^d}{(\Theta (\sigma)\bar{\mu}')^d} 
    {\log\frac{5\sqrt{d}\lambda'}{\e}}.
  $$
$ n_{\sigma}(\e)$ 
  is also a bound on the time to compute the full cells.
\label{lem:complexity-a-center}
\end{lemma}

\subsection{Construction of $\rdelo(L',W)$}
\label{sec:fastalgo}


By Lemma~\ref{lem:rdel.props}, all the simplices incident to a vertex $p'$ of 
$\del^{2\e}(L')$ are contained in $N(p') = L' \cap B(p', 2 \lambda' +4\e)$, and it follows 
from the fact that $L'$ is a $(\lambda', \bar{\mu}')$-net that 
$$
  |N(p')| \leq \frac{2^{O(d)}}{(\bar{\mu}')^{d}}.
$$
In the first step of the algorithm, we compute, for each $p'\in L'$, the set $N(p')$, 
and the set of $d$-simplices 
$$
  C_{d}(p') = \left\{ \sigma = \{p'\} \cup \tilde{\sigma}:~\mbox{$|\tilde{\sigma}| = d$ 
  and $\tilde{\sigma} \subset N(p') \setminus \{ p' \} $} \right\}.
$$
Observe that 
$$
  |C_{d}(p')| = \binom{|N(p')|}{d} = \frac{2^{O(d^{2})}}{(\bar{\mu}')^{d^{2}}}.
$$
We then extract from $C_d(p')$ a 
subset $WC_{d}(p')$ of simplices that have a full-leaf-point that is a $2\e$-Delaunay 
center, and have a number of full cells less than or equal to 
$$
n_0(\e) \stackrel{\rm def}{=}
U_d \left( \frac{4d}{\Theta_{0}\bar{\mu}'} \right)^d \log\frac{5\sqrt{d}\lambda'}{\e}.
$$
This is done by applying the algorithm of Section~\ref{sec:computing-acenters} with a twist.  
As soon as a $d$-simplex appears to have more than $n_0(\e)$ full cells, we stop considering 
that simplex.  The union of the sets $WC_d(p')$ for all $p'\in L'$ is a  subcomplex of $\del^{2\e}(L')$ called $\rdelo(L',W)$.
It contains every $d$-simplex $\sigma$ in $\del^{2\e}(L',W)$ that has a 
$2\e$-Delaunay center in $W$ at a distance less than $\e$ from its actual circumcenter, and 
satisfies the thickness criterion $\Theta(\sigma) \geq \Theta_{0}$.  Note, however, that we 
do not claim that every simplex that has at most $n_0(\e)$ full cells is $\Theta_{0}$-thick.
Algorithm~\ref{alg:rdelo} 
describes the construction of $\rdelo(L,W)$.
As noted above,  $|N(p')|=\frac{2^{O(d)}}{(\bar{\mu}')^d}$ for any $p'\in L'$, and all 
the $N(p')$ can be computed in $O(|L'|^{2})$ time by a brute fore method.
But assuming we have access to ``universal hash functions'' then we can use ``grid method'' 
described in \cite[Chapter~1]{Har-peled:GAA} with the sparsity condition of $L$ to get the 
complexity down to 
$$
  \frac{2^{O(d)}|L'|}{(\bar{\mu}')^{d}}.
$$
%
Using the facts that $\lambda' < 2 \lambda$ and $\bar{\mu}' \geq \frac{\bar{\mu}}{3}$,
see Lemma~\ref{lem:perturbed-net-parameters}, we conclude
that the total complexity of the algorithm is
$$
  O\left( \frac{|L|}{\Theta^{d}_{0}
      \,\bar{\mu}^{d^{2}+d}}\,\log \frac{\lambda}{\e}\right)
$$ 
and is therefore sublinear in $|W|$.
The constant in $O$ depends only on $d$.

\begin{algorithm}[ht]
  \caption{Construction of $\rdelo(L',W)$}
  \label{alg:rdelo}
  \begin{algorithmic}
    \STATE{\bf input:}\quad  a $(\lambda',\bar{\mu}')$-net $L'$, a grid
    $W$ of resolution $\e$, and $\Theta_{0}$ (defined in
    Eq.~\ref{eq:defn.ThetaO} and \ref{eq:defn.delta}))  
    \STATE{ $\Sigma := \emptyset$}
    \FOR{each $p'\in L'$} 
    \STATE{ compute the sets $N(p')= L'\cap B(p', 2\lambda' + 4\e)$ and $C_{d}(p')$}
    \ENDFOR
    \FOR{each $p' \in L'$}
    \FOR{each $d$-simplex $\sigma \in C_{d}(p')$} 
    \STATE // find the simplices in $C_{d}(p')$ that are in $WC_{d}(p')$
    \IF{$\sigma$ has a full-leaf-point that is a $2\e$-Delaunay center  and $n_{\sigma}(\e) \leq n_0(\e)$} 
        \STATE{add $\sigma$ to $\Sigma$}
     \ENDIF
    \ENDFOR
    \ENDFOR 
   \STATE{\bf output:}\quad $\Sigma$ the set of $d$-simplices of $\rdelo(L',W)$
  \end{algorithmic} 
\end{algorithm}


\subsection{Correctness of the algorithm}
\label{sec:sublin.correct}

We will need the following lemma which is an analog of Lemma~\ref{lemma-wit=del}.
The lemma also bounds  $\Theta_0$. Its proof follows directly from Lemma~\ref{thm:prot.thick}, 
and the observation that any simplex with a protected circumcentre is a Delaunay simplex.
\begin{lemma}
  \label{lem:rdel=del}
  Suppose that the $d$-simplices in $\del(L')$ are $\delta$-protected at their circumcenters, 
  with $\delta = \bar{\delta}\lambda'$.  
  If $0< \Theta_0 \leq \frac{\bar{\delta}\bar{\mu}}{24d}$,
  then $\del(L') \subseteq \rdelo(L',W)$ and if, in addition, every $d$-simplex of $\rdelo(L',W)$ 
  has a protected circumcenter, then $\rdelo(L',W)=\del(L')$.
\end{lemma}

We first show that Algorithm~\ref{alg:rdel} terminates if we
deactivate the call to procedure  \texttt{check()}. 
As discussed after Lemma~\ref{lem:rdel.props}, the analysis of Section~\ref{sec:lll.analysis} implies that the
perturbations of Algorithm~\ref{alg:rdel} can be expected to
produce a point set $L'$ for which all the $d$-simplices in
$\del^{2\e}(L',\T^d)$ have a $\delta$-protected circumcentre.  Since
this complex includes both $\del(L')$ \emph{and} $\rdelo(L',W)$,
Lemma~\ref{lem:rdel=del} shows that we can expect the 
algorithm to terminate with the condition that $\rdelo(L',W)$ has good
links.
%

We now examine procedure \texttt{check()} and show that it does not
affect the termination guarantee.
%
%
By Lemma~\ref{thm:prot.thick}, if $\del(L')$ is $\delta$-protected, 
then any $\sigma \in \del(L')$ satisfies 
\begin{equation}
  \label{eq:defn.ThetaO}
  \Theta (\sigma ) \geq  \frac{\bar{\delta}\bar{\mu}'}{8d}
  \geq \frac{\bar{\delta}\bar{\mu}}{24d}
  \stackrel{\text{def}}{=} \Theta_0.  
\end{equation}
Consider now $\sigma \in \rdelo(L',W)$.  Since the full leaves of the 
pyramid data-structure for $\sigma$ are composed entirely of 
$2\e$-centres at a distance less than $4\sqrt{d}\lambda'$ from any 
vertex of $\sigma$, Lemma~\ref{lem:a.del.ctrs} implies that, if 
$\sigma$ is $\Theta_0$-thick, then
\begin{equation*}
  \|x - c_{\sigma}\| \leq \frac{8\sqrt{d}\e}{\Theta_{0}\bar{\mu}'}.
\end{equation*}
This means that we can restrict our definition of $\rdelo(L',W)$ to 
include only simplices for which the set of full leaves has diameter 
less than $\frac{16\sqrt{d}\e}{\Theta_{0}\bar{\mu}'}$.
Further, we observe that if $\sigma$ is $\delta$-protected at its 
circumcentre, then it will have a $(\delta-2\e)$-protected full-leaf-point; 
this follows from the triangle inequality. The \texttt{check()} procedure 
described in Algorithm~\ref{alg:check} ensures that all the simplices 
in $\rdelo(L',W)$ have these two properties. It follows from the
discussion above that activating procedure \texttt{check()} does
not affect the termination guarantee.
\begin{algorithm}[ht]
  \caption{procedure \texttt{check}$(p')$}
  \label{alg:check}
  \begin{algorithmic}
    \IF{ all $d$-simplices $\sigma \in \str(p'; \rdelo(L',W))$
      satisfy
      \STATE 1. The diameter of the full leaves is at most 
      $\frac{16\sqrt{d}\e}{\Theta_{0}\bar{\mu}'}$.
      \STATE 2. There is a $(\delta - 2\e)$-protected full-leaf-point
    }
    \STATE{ {\bf check}$(p') = \mathrm{TRUE}$}
    \ELSE{ \STATE{\bf check}$(p') = \mathrm{FALSE}$}
    \ENDIF
  \end{algorithmic}
\end{algorithm}

The fact that the algorithm terminates yields
$\rdelo(L',W)$ with good links.
In order to apply Lemma~\ref{lem:rdel=del} to guarantee that
$\rdelo(L',W)=\del (L')$, we need to guarantee that  the simplices of
$\rdelo(L',W)$ are protected.
The analysis of the Moser-Tardos algorithm in Section~\ref{sec:lll.analysis} provides a possible $\delta$-protection with
\begin{equation}
  \label{eq:defn.delta}
  \delta = J\rho,
\end{equation}
where $J$ is defined in Eq.~\eqref{eq:condition-parameters-relax-1}. However, we cannot guarantee that this protection is actually obtained. 
The following lemma shows that if the perturbation $\rho$ (and consequently $\delta$) is sufficiently large, then the output point set will be $\delta^*$-protected for a positive $\delta^*$.
\begin{lemma}
  	\label{lem:defn.delta.star}
        The $d$-simplices in $\rdelo(L',W)$ produced by Algorithm~\ref{alg:rdel} are $\delta^*$-protected, with
  	\begin{equation*}
	  \delta^* = \delta
          - \left(\frac{34\sqrt{d}}{\Theta_0\bar{\mu}'}\right) \e,
  	\end{equation*}
        where $\delta = J\rho$.
\end{lemma}
\begin{proof}[of Lemma~\ref{lem:defn.delta.star}]
  	Let $\sigma \in \rdelo(L',W)$ be a $d$-simplex produced by Algorithm~\ref{alg:rdel}, and 
	let $w$ be the full-leaf-point for $\sigma$ found by the \texttt{check()} procedure. Since 
	$c_{\sigma}$ is located in a full leaf-node, the restriction on the diameter of the leaf nodes implies
  	\begin{equation*}
    		\norm{c_{\sigma} - w} < \frac{16\sqrt{d}\e}{\Theta_0\bar{\mu}'}.
  	\end{equation*}
  	Since $w$ is a $(\delta-2\e)$ protected $2\e$-centre, we have that for all $p \in \sigma$ 
	and $q \in L' \setminus \sigma$:
  	\begin{equation*}
    		\begin{split}
      			\norm{q - c_{\sigma}} &\geq \norm{q - w} - \norm{w - c_{\sigma}}\\
      			&> \norm{p-w} + (\delta - 2\e) - \frac{16\sqrt{d}\e}{\Theta_0\bar{\mu}'}\\
      			&> \norm{p-c_{\sigma}} + (\delta - 2\e)
      			- \frac{32\sqrt{d}\e}{\Theta_0\bar{\mu}'}\\
      			&> \norm{p-c_{\sigma}}
      			+ \left(\delta - \frac{34\sqrt{d}\e}{\Theta_0\bar{\mu}'}\right).
    		\end{split}
  	\end{equation*}
\end{proof}

In order to have
$\delta^* > 0$, we need a lower
bound on $\delta$, and hence on the minimal perturbation radius through 
$\delta = J\rho$. Therefore
we require:
\begin{equation*}
  \frac{J\mu}{4} \geq \delta > \frac{34\sqrt{d}\e}{\Theta_0\bar{\mu}'}
\end{equation*}
(compare with \eqref{eq:condition-parameters-relax}). Writing $\bar{\delta} = \frac{\delta}{\lambda'}$ and $\Theta_{0} = \frac{\bar{\delta} \bar{\mu}}{24d}$, and using $\lambda'\geq \lambda$,
we obtain the conditions under which Algorithm~\ref{alg:rdel} is guaranteed to produce a $\delta^*$-protected Delaunay triangulation:
\begin{equation*}
  \frac{J\bar{\mu}}{4}\geq \bar{\delta} > \frac{2448 d^{\frac{3}{2}}}{\bar{\delta}\bar{\mu}^2}\frac{\e}{\lambda}.
\end{equation*}
The right-hand inequality is satisfied provided 
$$
  \bar{\delta} \geq \frac{50 d^{\frac{3}{4}}}{\bar{\mu}}\, \sqrt{\frac{\e}{\lambda}}.
$$
We have proved
\begin{theorem}
  \label{thm:rdel.guarantees}
  If $\bar{\rho}\leq \bar{\mu}/4$ and 
  $\bar{\rho} = \Omega\left(\frac{1}{J\bar{\mu}}\,  \sqrt{ \frac{\e}{\lambda}}\right)$ 
  (with $J$ defined in Eq.~\eqref{eq:condition-parameters-relax-1}), Algorithm~\ref{alg:rdel} 
  terminates and outputs the Delaunay triangulation of $L'$. The Delaunay $d$-simplices are 
  $\delta^*$-protected, as defined in Lemma~\ref{lem:defn.delta.star}, and consequently satisfy 
  a thickness bound of
  $$
    \Theta(\sigma) \geq \frac{\bar{\delta}^*(\bar{\mu}/3 +
    \bar{\delta}^*)}{8d} \stackrel{\rm def}{=}\Theta_*.
  $$

  The complexity of the algorithm is 
  $$
    O\left(\frac{|L|}{\Theta_{*}^{d}\, \bar{\mu}^{d^2+d}}\log \frac{\lambda}{\e}\right).
  $$
  The constants in $\Omega$ and $O$ depend only on $d$. 
%
\end{theorem}

\section{Boundaries and Phantom points}
\label{app-boundary-issues}

We work on $\T^d$ because our algorithm is not designed to compute 
a triangulation with boundaries. If we are given a finite set 
$P \subset \R^d$, we scale it into the unit box, and then extend it 
periodically. In effect, by working on the torus, we are adding 
``phantom points'' to the given point set, in order to avoid the 
boundary problem.

The Delaunay triangulation that we compute will coincide with the 
Delaunay triangulation of the convex hull of $P \subset \R^d$ only 
for points sufficiently far from the boundary of $\conv(P)$. Specifically, 
suppose $P$ is a $\gamma$-sample\footnote{If $D \subset \R^d$, we 
say that $P$ is a {$\gamma$-sample for $D$} if any point in $D$ is 
a distance less than $\gamma$ from a point in $P$.} of $\conv(P)$, 
and let $L = P \cup Q$ where $Q$ is a set of phantom points (e.g., 
periodic copies of $P$) such that $Q \cap \conv(P) = \emptyset$. 
Let $\del_|(P)$ be the Delaunay triangulation of $P$ restricted to
\begin{equation*}
	D_{\gamma}(P) = \left\{ x \in \conv(P) \, \big| \, d(x,\partial(\conv(P))) \geq \gamma \right\},
\end{equation*}
i.e., $\del_|(P)$ is the set of simplices in $\del(P)$ that have a Delaunay 
ball centered in $D_{\gamma}(P)$. Then $\del_|(P) \subseteq \del(L)$. Indeed, 
by construction, a Delaunay ball $B$ for $\sigma \in \del_|(P)$ must be 
contained in $\conv(P)$, and since $Q \cap \conv(P) = \emptyset$, it follows 
that $B$ is also a Delaunay ball in $L$.

The extended set $L \subset \R^d$ will be a $\lambda$-sample for the unit 
hypercube, and we observe that this sampling radius $\lambda$ may be larger 
than $\gamma$ in general. But that is not relevant to the above observation. 
The quantity $\gamma$ is only relevant for describing $\del_|(P)$, but this 
complex is not considered by the algorithm: the algorithm is dependent only 
on the parameters $\lambda$ and $\bar{\mu}$ that describe $L$.

Our demand that $\lambda \leq 1/4$ is the price we pay for the convenience 
of working on the flat torus, where we identify the phantom points $Q$ as 
copies of points of $P$. This allows us to describe the algorithm by 
considering only the given set of points in the unit hypercube. 

We now briefly describe an alternative approach which involves remaining 
in $\R^d$ and explicitly adding additional points to the given point set 
$P$ so that the boundary of the whole point set does not interfere with 
our guarantees on the simplices whose vertices are in $P$.

If we are interested in using witnesses to compute the Delaunay complex on 
a finite vertex set in $\R^d$, we need to confine our considerations to a 
bounded domain $D \subset \R^d$. But this is not possible if a Delaunay 
simplex can have an arbitrarily large circumradius.  

Assuming only that the finite set of distinct points $P$ contains a 
non-degenerate $d$-simplex, we can introduce a set $Q$ of 
\defn{phantom vertices} such that the set $P \cup Q$ is a 
$(\lambda,\bar{\mu})$-net for some ball $D \subset \R^d$ with finite radius.
Without loss of generality (scaling if necessary), we can assume that $P$ is 
contained in the unit ball centred at the origin. Let $Q$ be the set of $2d$ 
points defined by the intersection of the coordinate axes with the boundary 
of an axis-aligned cube centred at the origin, and whose sides are of length 
$4\sqrt{d}$.

%
The convex hull of $Q$ is the cross-polytope, the dual of the cube. Each 
facet is an equilateral simplex with circumradius $2\sqrt{(d-1)}$. Any point 
of $P$ is at least unit distance from the affine hull of any boundary facet, 
and an elementary calculation shows that the circumradius of any simplex in 
$\del(P \cup Q)$ must be less than $2d - \frac{3}{2}$. It follows then that 
regardless of the dimension $d$, the points of $P \cup Q$, as well as the 
circumcentres of all the simplices in $\del(P \cup Q)$, are contained in a 
ball $D$ of radius $2d$ centred at the origin. 

Any Delaunay ball for a facet of $\conv(Q)$ that is centred at a distance greater 
than $2d - 1/2$ from the origin is protected in $\del(P\cup Q)$, and being 
equilateral, these facets have maximal thickness. These observations imply 
that the guarantees of our algorithms can be shown to apply if we take $L=P \cup Q$, 
but only perform perturbations and link tests on the points in $P$. For the 
analysis in this case, the lemmas which reference $\str^2(p)$ would need to 
be modified. Specifically, by using Proposition~\ref{prop:inherit.protection} 
directly, instead of the simplified Lemma~\ref{lem:protection-inheritance}, we 
can still assume the protection of lower-dimensional simplices as is required 
for the Witness algorithm of Section~\ref{sec:compute-del-via-wit}. Similarly, 
the thickness of Delaunay simplices that have some vertices in $Q$ can be 
ensured by using the inherent thickness and protection of the simplices that 
have all vertices in $Q$ as a substitute for the protection of neighbouring 
Delaunay $d$-simplices required by Lemma~\ref{thm:prot.thick}.

Therefore, in principle the algorithm can be adapted to work with a finite set, 
rather than a periodic one, but there is no clear advantage. The algorithm, and 
its analysis would become more complicated. Also, the sampling radius $\lambda$ 
will be very large in this setting, implying that the sparsity parameter 
$\bar{\mu}$ would be very small, adversely affecting the running time. The subcomplex 
of $\del(P\cup Q)$ that is guaranteed to belong to $\del(P)$, is exactly the 
complex $\del_|(P)$ mentioned above.

If $P \subset \R^d$ is a $\gamma$-sample for $\conv(P)$, and we could identify 
the points $\tilde{P} \subset P$ that are at a distance at least $4\gamma$ from 
$\partial\conv(P)$, then we could perform the algorithm by only applying the 
link test on the points of $\tilde{P}$ (the deep interior points 
\cite{bdg-stabdt-2014}). In this case, the algorithm will terminate, but only 
the simplices incident to points of $\tilde{P}$ would be guaranteed to be Delaunay, 
and this is generally a smaller subcomplex than $\del_|(P)$.

\section{Discussion}

In earlier work~\cite{boissonnat:hal-00806107} we described a perturbation algorithm 
to produce a $\delta$-protected point set for Delaunay triangulations. By employing 
the Moser-Tardos algorithmic framework, as we have done here, we obtain a much simpler 
analysis for the same result. Indeed, it is precisely the construction of a $\delta$-protected 
point set that allows us, in Section~\ref{sec:algo}, to produce a witness complex that 
coincides with the Delaunay triangulation. In the previous work, each point is only 
perturbed once, whereas here points are only perturbed as needed, but the Moser-Tardos 
result ensures that only a linear number of perturbations are made in total. As 
indicated in \eqref{eq:condition-parameters-relax-1}, the new analysis yields protection 
(and therefore thickness) of the order $(1/2)^{O(d^2)}$, improving on the $(1/2)^{O(d^3)}$ 
obtained in~\cite{boissonnat:hal-00806107}.


The techniques described in this paper can
also be applied to construct {\em weighted Delaunay triangulations} and 
{\em Delaunay triangulations on manifolds}
as will be reported elsewhere.

\section*{Acknowledgement}
 
This research has been partially supported by the 7th Framework
Programme for Research of the European Commission, under FET-Open
grant number 255827 (CGL Computational Geometry Learning). Partial 
support has also been provided by the Advanced Grant of the European 
Research Council GUDHI (Geometric Understanding in Higher Dimensions).

Arijit Ghosh is supported by the Indo-German Max Planck Center for Computer Science (IMPECS).

The authors would like to thank Steve Y. Oudot and Mariette Yvinec
for discussions on some of the concepts developed in this work.

\appendix

%
\newcommand{\splxs}{\sigma}

\newcommand{\splxt}{\tau}
\newcommand{\splxjoin}[2]{{#1}*{#2}}

\newcommand{\bigo}[1]{\mathcal{O}(#1)}

\newcommand{\samconst}{\lambda}
\newcommand{\sparseconst}{\bar{\mu}}
\renewcommand{\sphere}[2]{S(#1,#2)} 

\newcommand{\glift}{\phi}
\newcommand{\lift}[1]{\glift(#1)}
\newcommand{\vorcell}[1]{\mathrm{Vor}(#1)}

\newcommand{\dotprod}[2]{#1\cdot#2}

\newcommand{\Lemref}[1]{Lemma~\ref{#1}}

\section{Inheritance of protection}
\label{sec:protect.all}

If $\pts \subset \R^d$ is $\delta$-protected, then by definition every $d$-simplex 
$\splxs \in \del(L)$ has a $\delta$-protected Delaunay ball. In this section we quantify 
the amount of protection this guarantees for simplices of lower dimension. 


We will demonstrate:
\begin{proposition}
  \label{prop:inherit.protection}
  Suppose $L \subset \R^d$ is a locally finite set, and $\sigma \in \del(L)$ is a non-degenerate 
  $k$-simplex whose Delaunay balls all have radius less than $\lambda$.  If $\vor(\sigma)$ 
  is bounded, and every $d$-simplex in $\del(L)$ that has $\sigma$ as a face has circumradius 
  at least $\frac{1}{2}\bar{\mu}\lambda$, and is $\delta$-protected, with $\delta=\bar{\delta}\lambda$, 
  and $\bar{\delta} \leq \bar{\mu} \leq 2$, then $\sigma$ is 
  $$
    \frac{(\bar{\mu}+\bar{\delta})\delta}{4(d-k+1)}\mbox{-protected}.
  $$
  Specifically, letting $j=d-k$, there exists $j+1$ $d$-simplices 
  $\{\sigma_{0}, \, \dots, \, \sigma_{j}\} \subseteq \del(L)$, each having $\sigma$ as a face, and 
  $\sigma$ is $\frac{(\bar{\mu} + \bar{\delta})\delta}{4(j+1)}$-protected at 
  $$
    c^* = \frac{1}{j+1} \sum_{i=0}^{j} c_{\sigma_{i}}.
  $$
\end{proposition}
If $k=0$, then $\sigma = \{p\}$, and it is easy to see directly that $\sigma$ must be 
$\delta$-protected: if $q \in L$ and $q \neq p$, then there is a $d$-simplex 
$\sigma^d \in \del(L)$ such that $p \in \sigma^d$ and $q \not \in \sigma^d$. Since 
$\sigma^d$ is $\delta$-protected, we have $\norm{p-q}>\delta$, and it follows that 
the trivial Delaunay ball for $p$ is $\delta$-protected. Therefore the smallest lower 
bound on the protection of lower dimensional simplices happens when $k=1$.
With this observation, Lemma~\ref{lem:protection-inheritance} follows immediately from 
Proposition~\ref{prop:inherit.protection}, and the fact that the protection of 
the $d$-simplices in $\str(p)$ implies there are no degenerate 
simplices~\cite[Lemma~3.5]{bdg-stabdt-2014}.

A standard exercise establishes that $L$ is generic if and only if the Voronoi cell 
associated to any $k$-simplex has dimension $j=d-k$. Rather than requiring $L$ to be 
generic, Proposition~\ref{prop:inherit.protection} makes the more local assumption 
that $\sigma$ is non-degenerate. In this setting, this is sufficient to ensure that 
$\vor(\sigma)$ has dimension $d-k$. To see this, first observe that a short argument 
reveals that since all the $d$-simplices that contain $\sigma$ are $\delta$-protected, 
none of them can be degenerate. Then another short argument shows that all the faces 
of a non-degenerate protected $d$-simplex must have Voronoi cells of the correct 
dimension.

It will be convenient to work with squared distances and the ``lifting map'', defined 
below, and so we introduce a modified notion of protection, called 
\defn{power-protection}, to accommodate this.

\subsection{Power-protection, and its relation to protection}

We say that a Delaunay ball $B = B(c,r)$ for $\splxs$ is
$\delta^2$-\defn{power-protected} if 
$$
  \| c-q\| ^2 - r^2 > \delta^2,\; \forall \, q \in \pts \setminus \splxs.
$$
The following result is a special case of a result on power-protection
proved in~\cite[Lemma~8]{2014arXiv1410.7012B}.
\begin{lemma}
  \label{lem:inherit.pwr.protection}
  Suppose $L \subset \R^d$ is a locally finite set, and $\sigma \in \del(L)$ 
  is a non-degenerate $k$-simplex.  If $\vor(\sigma)$ is bounded, and every 
  $d$-simplex in $\del(L)$ that has $\sigma$ as a face is 
  $\delta^2$-power-protected, then $\sigma$ is 
  $$
    \frac{\delta^2}{(d-k+1)}\mbox{-power-protected.}
  $$
  
  Specifically, letting $j=d-k$, there exists $j+1$ $d$-simplices 
  $\{\sigma_{0}, \, \dots, \, \sigma_{j}\} \subseteq \del(L)$, each having 
  $\sigma$ as a face, and $\sigma$ is 
  $\frac{\delta^2}{j+1}$-power-protected at 
  $$
    c^* = \frac{1}{j+1} \sum_{i=0}^{j} c_{\sigma_{i}}.
  $$
\end{lemma}

Protection implies power-protection.
Specifically, Let $B = B(c,r)$ be a Delaunay ball for $\splxs$,
and let $R = \min_{q\in  L\setminus \splxs} \| c-q\|$. If $B$ is
$\delta$-protected, then $R-r > \delta$. Assume that the circumradius of
$\splxs$ is at least $\frac{1}{2}\bar{\mu}\lambda$.
Then
\begin{equation*}
  r \geq \frac{1}{2}\sparseconst \samconst, 
  \quad \text{and} \quad
  R > r + \delta \geq \frac{1}{2}\sparseconst \samconst + \delta.
\end{equation*}
Thus, using $\delta = \bar{\delta}\lambda$, we find
\begin{equation*}
  R^2 - r^2 = (R+r)(R-r) > (\sparseconst \samconst + \delta)(R-r) > 
  \sparseconst \samconst \delta + \delta^2 = (\sparseconst\bar{\delta} 
  + \bar{\delta}^2)\lambda^2,
\end{equation*}
which implies:
\begin{lemma}
  \label{lem:prot.to.power}
  Suppose $\splxs \in \del (L))$ has circumradius at least $\bar{\mu}\lambda$. 
  If $\splxs$ is $\delta$-protected, then it is 
  $((\sparseconst\bar{\delta} + \bar{\delta}^2)\lambda^2)$-power-protected.
\end{lemma}

Going the other way, assume that $\splxs$ is
$\delta^2$-power-protected, so that $R^2 -r^2 = \delta^2$.  Assume further, that 
$r < \lambda$, and that $\delta^2 \leq 8\lambda^2$. Then $\sigma$ is 
$\delta'$-protected, where
\begin{equation*}
  \delta' = (R-r) = \frac{1}{(R+r)}(R^2-r^2) > \frac{\delta^2}{2\lambda + \delta'}.
\end{equation*}
If $\delta' \leq 2\lambda$, then
\begin{equation}
  \label{eq:prot.bnd}
  \delta' \geq \frac{\delta^2}{4\lambda},
\end{equation}
and by assumption we have $2\lambda \geq \frac{\delta^2}{4\lambda}$ anyway, so 
Inequality~\eqref{eq:prot.bnd} is always true. We have shown:
\begin{lemma}
  \label{lem:power.to.prot}
  Suppose $B = B(c,r)$ is a $\delta^2$-power-protected Delaunay ball for 
  $\sigma \in \del(\pts)$. If $r < \lambda$, and $\delta^2 \leq 8\lambda^2$, then 
  $B$ is a $\delta'$-protected Delaunay ball for $\sigma$, where 
  $$
    \delta' = \frac{\delta^2}{4\lambda}.
  $$
\end{lemma}
As the notation suggests, we can employ both \Lemref{lem:prot.to.power}, and 
\Lemref{lem:power.to.prot} whenever $L$ is a $(\lambda,\bar{\mu})$-net. 

\begin{proof}[of Proposition~\ref{prop:inherit.protection}]
  Given the $\delta$-protection and lower bound on the radius of the $d$-simplices that are 
  co-faces of $\sigma$, Lemma~\ref{lem:prot.to.power} ensures that these simplices are 
  $((\bar{\mu}+\bar{\delta})\delta\lambda)$-power-protected. Then Lemma~\ref{lem:inherit.pwr.protection} 
  ensures that $\sigma$ is
  $\frac{(\bar{\mu}+\bar{\delta})\delta\lambda}{d-k+1}$-power-protected at $c^*$.

  We observe that the condition $\bar{\delta} \leq \bar{\mu} \leq 2$, implies that 
  $(\bar{\mu}\bar{\delta} +\bar{\delta}^2)\lambda^2 \leq 8\lambda^2$, and so
  the bound on the radius of the Delaunay ball for $\sigma$ centred at $c^*$ allows us to 
  employ Lemma~\ref{lem:power.to.prot} to find that $\sigma$ is
  $\frac{(\bar{\mu}+\bar{\delta})\delta}{4(d-k+1)}$-protected at $c^*$.
\end{proof}

\bibliographystyle{alpha}

\bibliography{biblio}



\end{document}